\PassOptionsToPackage{unicode}{hyperref}
\PassOptionsToPackage{hyphens}{url}
\documentclass[
]{article}
\usepackage{amsmath,amssymb}
\usepackage{iftex}
\ifPDFTeX
  \usepackage[T1]{fontenc}
  \usepackage[utf8]{inputenc}
  \usepackage{textcomp} 
\else 
  \usepackage{unicode-math} 
  \defaultfontfeatures{Scale=MatchLowercase}
  \defaultfontfeatures[\rmfamily]{Ligatures=TeX,Scale=1}
\fi
\usepackage{lmodern}
\ifPDFTeX\else
\fi
\IfFileExists{upquote.sty}{\usepackage{upquote}}{}
\IfFileExists{microtype.sty}{
  \usepackage[]{microtype}
  \UseMicrotypeSet[protrusion]{basicmath} 
}{}
\makeatletter
\@ifundefined{KOMAClassName}{
  \IfFileExists{parskip.sty}{%
    \usepackage{parskip}
  }{
    \setlength{\parindent}{0pt}
    \setlength{\parskip}{6pt plus 2pt minus 1pt}}
}{
  \KOMAoptions{parskip=half}}
\makeatother
\usepackage{xcolor}
\usepackage[margin=0.75in]{geometry}
\usepackage{graphicx}
\usepackage{caption}
\usepackage{float}
\usepackage{placeins}
\usepackage{pdflscape}
\graphicspath{{figures/}}
\usepackage{longtable,booktabs,array}
\usepackage{calc} 
\usepackage{etoolbox}
\makeatletter
\patchcmd\longtable{\par}{\if@noskipsec\mbox{}\fi\par}{}{}
\makeatother
\IfFileExists{footnotehyper.sty}{\usepackage{footnotehyper}}{\usepackage{footnote}}
\makesavenoteenv{longtable}
\ifLuaTeX
  \usepackage{selnolig}  
\fi
\usepackage{bookmark}
\IfFileExists{xurl.sty}{\usepackage{xurl}}{} 
\hypersetup{
  hidelinks,
  pdftitle={Finch: Toxicity Dose Response Curve Prediction of Chemical Compounds and Mixtures},
  pdfcreator={LaTeX via pandoc}}

\author{}
\date{}

\begin{document}

\noindent\textbf{Article}

\begin{center}
{\LARGE\bfseries Finch: Toxicity Dose Response Curve Prediction of Chemical Compounds and Mixtures\par}
\vspace{0.8em}
Abdullah Shouaib \textsuperscript{1}, John Zapanta \textsuperscript{1}, Sean P. Davern \textsuperscript{1}, Samuel Dixon \textsuperscript{1}, Zachary R. Stromberg \textsuperscript{1}, Becky Hess \textsuperscript{1}, Sydney Schwartz \textsuperscript{1}, and C Mark Maupin \textsuperscript{1,}*
\end{center}

\textsuperscript{1} Pacific Northwest National Laboratory, Richland, Washington 99354, USA; \href{mailto:Abdullah.shouaib@pnnl.gov}{\nolinkurl{Abdullah.shouaib@pnnl.gov}} (A.S.); \href{mailto:john.zapanta@pnnl.gov}{\nolinkurl{john.zapanta@pnnl.gov}} (J.Z.); \href{mailto:sean.davern@pnnl.gov}{\nolinkurl{sean.davern@pnnl.gov}} (S.P.D.); \href{mailto:Samuel.dixon@pnnl.gov}{\nolinkurl{Samuel.dixon@pnnl.gov}} (S.D.); \href{mailto:Zachary.stromberg@pnnl.gov}{\nolinkurl{Zachary.stromberg@pnnl.gov}} (Z.R.S.); \href{mailto:Becky.Hess@pnnl.gov}{\nolinkurl{Becky.Hess@pnnl.gov}} (B.H.); \href{mailto:schwartz.sydneyc@gmail.com}{\nolinkurl{schwartz.sydneyc@gmail.com}} (S.S.); \href{mailto:mark.maupin@pnnl.gov}{\nolinkurl{mark.maupin@pnnl.gov}} (C.M.M)

\textbf{*} Correspondence: \href{mailto:mark.maupin@pnnl.gov}{\nolinkurl{mark.maupin@pnnl.gov}} ; Tel.: +1 509-375-3738

\textbf{Abstract}

The need to holistically view human interactions with chemical mixtures is driving a paradigm shift in the chemical risk assessment field that necessitates product/mixture over single compound testing, elimination of animal testing, and new modeling approaches to understand mixture activity profiles. Unfortunately, most computational models primarily focus on the analysis and estimation of single chemical species with very few viable mixture models that go beyond conventional modeling approaches. Conventional mixture modeling approaches, such as concentration addition (CA) and independent action (IA), are limited by their ability to handle multiple Modes of Action (MoA), and often overlook synergistic/antagonistic effects. Therefore, a fast and efficient model for mixture prediction that goes beyond conventional approaches would be an asset to regulatory entities that are concerned with safety profiles. Finch is a novel workflow that addresses these challenges by not only providing a state-of-the-are molecular descriptor-based framework but also leveraging deep learning (DL) embeddings in multi-task quantitative structure-activity relationship (QSAR) models for enhanced chemical exposure prediction. The use of DL embeddings maximizes information preservation in the latent space from a vast number of inputs, including molecular descriptors, physiochemical properties, and large language model (LLM) embeddings derived from a Simplified Molecular Input Line Entry System (SMILES) input. DL embeddings are employed to distill critical features and preserve information into a latent space, enhancing the predictive capability of subsequent machine learning models. The multi-task learning aspect of Finch is particularly advantageous as it allows for the simultaneous optimization of multiple loss functions. This approach is beneficial as it uses all the available data across the different tasks to learn generalized representations, which can aid in effectively capturing complex ingredient interactions within mixtures.


\section*{1. Introduction}

Predicting the dose--response behavior of mixtures is a critical challenge in toxicology, driven by the multitude of ways in which chemical agents can interact within biological systems {[}1-3{]}. Unlike single-agent exposures, mixtures often exhibit non-linear interactions, including antagonistic, synergistic, or potentiation effects {[}4, 5{]}. Anticipating these interactions accurately is vital for safeguarding public health and environmental safety, as well as for guiding regulatory decisions and product development.

Traditional toxicological testing methods, which rely extensively on \emph{in vivo} and/or \emph{in vitro} experiments, can be time-consuming and costly for the vast array of potential chemical combinations. Fortunately, with recent advances in artificial intelligence (AI) and machine learning (ML), there are an increasing number of algorithms that offer a promising alternative through the integration of large-scale datasets and sophisticated modeling techniques {[}6{]}. For instance, deep learning architectures can uncover hierarchical features that capture intricate relationships among toxicological endpoints and chemical properties {[}7-9{]}.

While there are some successful studies in mixture toxicity models {[}10-17{]}, building robust predictive models in toxicology remains challenging. One critical hurdle involves limited data availability for complex mixtures, which can vary dramatically in composition, concentration, and endpoints of interest. Moreover, the sheer number of possible chemical combinations far exceeds the data generated by traditional toxicity testing, making it difficult to capture all relevant interactions. Experimental techniques, like designed experiments, that can enable quantification of single or multi-component interactions often require robotic equipment to be done robustly, are expensive, and limit labs that can perform such work.

In this context, transfer learning has emerged as a powerful approach, enabling models trained in one domain (e.g., single-chemical toxicity data) to be adapted for another (e.g., multi-component mixtures) {[}18-20{]}. By leveraging pre-established representations or learned parameters from extensive single-chemical datasets, transfer learning reduces the demand for large, mixture-specific training sets, which fundamentally addressing the data scarcity issue. Such methods often benefit from deep learning architectures or advanced machine learning algorithms that can detect subtle interactions among components. Beyond improving model performance and predictive power, these strategies can substantially reduce the number of experiments needed, dissecting complex mixture effects with greater efficiency. Ultimately, successfully implementing transfer learning in QSAR-based mixture toxicity modeling can expedite the identification of potentially hazardous combinations, guide targeted testing strategies, and support more informed risk assessment and regulatory decision-making.

In this work, we introduce Finch: an AI-powered toolbox designed to predict the physiological functions and hazard profiles of chemical mixtures. By integrating transfer approaches with large language model embeddings, Finch provides a systematic and flexible approach to modeling dose--response relationships in complex mixtures. Through analysis of multiple cytotoxicity endpoints, we demonstrate how Finch can serve as an invaluable tool for researchers, industry, and regulatory agencies in efficiently identifying and prioritizing chemical mixtures for further testing. Overall, Finch addresses a critical gap in computational toxicological modeling by consolidating transfer learning approach methodologies tailored to mixture toxicity.

\section*{2. Materials and Methods}

\subsection*{2.1 Data Collection and Processing}

\subsubsection*{2.1.1 Cytotoxicity of Mixtures Data}

Cytotoxicity data for individual compounds and mixtures were obtained from a previously established dataset {[}12, 16{]}, which contained information from the HBM4EU project. From the original dataset, 24 compounds (Supplemental Table 1) were selected, which included 9 heavy metals, 6 organophosphate flame retardants (OPFRS), 3 polyfluoroalkyl compounds (PFAS), and 6 phenols. In addition to the individual compounds, a total of 39 mixtures (Supplemental Table 2) were selected for this study. The dataset contained the Simplified Molecular Input Line Entry System (SMILES) for the compounds, mixture compositions, and toxicity measurements for all compounds and mixtures at different concentrations. The cytotoxicity values were capped at 100\% and subsequently normalized to a range of 0 to 1 using min-max scaling to facilitate model training. Concentration values were then converted from the originally reported value to Molar values (i.e., divided by 10\textsuperscript{6}), which was found to improve numerical stability during model training.

\subsubsection*{2.1.2 PubChem Bioassay Data}

The primary dataset was obtained from PubChem, consisting of bioassay data for HepG2 cell line toxicity at different exposure times (24 h and 40 h). The datasets (AID\_1224879\_datatable\_hepg2\_40h.csv and AID\_1224867\_datatable\_hepg2\_24h.csv) contain compound information including SMILES representations, activity outcomes, and various assay parameters for 9,524 compounds. Compounds were classified as either "Active" (toxic) or "Inactive" (non-toxic) based on their ``PUBCHEM\_ACTIVITY\_OUTCOME'' field. SMILES strings were extracted from the PubChem dataset and preprocessed for model input. Binary labels were created with 1 representing "Active" (toxic) compounds and 0 representing "Inactive" compounds. The dataset was split into training (80\%) and validation (20\%) sets using stratified sampling to maintain the class prevalence distribution.

\subsection*{2.2 ChemBERTa-2 Fine-Tuning}

We employed the ChemBERTa-2 model (available through the HuggingFace) to perform binary classification on compound cytotoxicity. All tasks, including data preprocessing, model fine-tuning, and embedding extraction, were conducted in Python (version 3.10) using pandas (version 1.4), numpy (version 1.24), PyTorch (version 2.5), and the Hugging Face Transformers (version 4.33) libraries. To prepare the chemical structures (i.e., SMILES strings) for the model, each sample was tokenized using ChemBERTa-2's custom tokenizer within the transformer's library. The tokenizer converts SMILES strings into a sequence of tokens suitable for the ChemBERTa-2 architecture, appending special tokens such as {[}CLS{]} and {[}SEP{]} consistent with the standard BERT-like encoder.

For the fine-tuning procedure, the PubChem Bioassay Data utilized where each input chemical was encoded, and the {[}CLS{]} output token was passed through a classification head to yield a probability score for cytotoxic (+) versus non-cytotoxic (\ensuremath{-}). Following training, the huggingface/transformers inference pipeline was used to encode each SMILES and extract the output {[}CLS{]} token. These embeddings, typically 384-dimensional for ChemBERTa-2, were then saved for subsequent downstream analyses which included clustering and additional machine learning model training. All experiments were performed on a DGX node with 8 2080-ti's to accelerate training.

The model was trained for 200 epochs using a batch size of 64, with the AdamW optimizer and a learning rate of 10\textsuperscript{-5}. We employed a linear learning rate scheduler with 500 warmup steps and a weight decay of 0.01. Model performance was evaluated using accuracy, F1-score, and the Receiver Operating Characteristic Area under the Curve (ROC-AUC) curve metrics on a held-out validation set comprising 20\% of the data.

\subsection*{2.3 Molecular Descriptor Generation}

Molecular descriptors for individual compounds from the previously obtained data set {[}12, 16{]} were generated using the R's Chemistry Development Kit (RCDK) {[}21{]}. RCDK created descriptors that capture various physicochemical properties of the compounds, including topological, geometric, and electronic features. Descriptors containing NAs or that had no variability were removed. The resulting 103 descriptors were then max-min normalized.

For chemical mixtures, a mole fraction weighted sum of individual molecular descriptors (MD) was used to calculate a formula molecular descriptor (FMD) for the mixture.

\[{FMD}_{i} = \sum_{n}^{n_{\max}}\left( x_{n} \times {MD}_{i,n} \right)\]

where FMD\textsubscript{i} is the i\textsuperscript{th} formula molecular descriptor for a given mixture, \ensuremath{\chi}\textsubscript{n} is the mole fraction of compound n in the selected mixture, and MD\textsubscript{i,n} is the i\textsuperscript{th} molecular descriptor for the n\textsuperscript{th} individual compound in the selected mixture. This approach yields a single composite descriptor vector that captures the aggregate molecular representation of the mixture. Once computed, these composite descriptors were used as input features to train random forest regression models.

\subsection*{2.4 Molecular Embedding Extraction}

For the pre-trained ChemBERTa model and the fine-tuned ChemBERTa model, the internal representations were leveraged to extract molecular embeddings that capture the chemical information relevant to toxicity prediction. These embeddings served as rich molecular descriptors for downstream machine learning tasks, particularly for the concentration-dependent toxicity prediction model.

To extract embeddings for individual compounds, we processed each SMILES string through the ChemBERTa and the fine-tuned ChemBERTa models and captured the output from the final hidden layer corresponding to the {[}CLS{]} token. This process was implemented using PyTorch\textquotesingle s no-gradient context to efficiently process batches of compounds. The resulting embeddings were 384-dimensional vectors that encapsulated the chemical information learned during the fine-tuning process.

For mixture modeling, we leveraged the same concept employed for molecular descriptors (FMDs) by computing a formula molecular embedding (FME) that is a weighted sum of the ChemBERTa {[}CLS{]} token embeddings. Specifically, for a mixture composed of multiple chemicals, we calculated each chemical\textquotesingle s embedding independently and then combined them in proportion to their respective mole fractions.

\[{FME}_{i} = \sum_{n}^{n_{\max}}\left( x_{n} \times \lbrack CLS\rbrack_{i,n} \right)\]

where FME\textsubscript{i} is the i\textsuperscript{th} formula molecular embedding for a given mixture, \ensuremath{\chi}\textsubscript{n} is the mole fraction of compound n in the selected mixture, and {[}CLS{]}\textsubscript{i,n} is the i\textsuperscript{th} molecular embedding (i.e., the i\textsuperscript{th} element of the {[}CLS{]} embedding) for the n\textsuperscript{th} individual compound in the selected mixture. This approach yields a single composite embedding vector that was subsequently used as input features to train random forest regression models.

\subsection*{2.5 Random Forest Models for Concentration-Dependent Toxicity Prediction}

Three Random Forest Models were created to compare the use of MDs and FMDs with the use of the pre-trained ChemBERTa and fine-tuned ChemBERTa MEs and FMEs for the purpose of predicting concentration-dependent cytotoxicity responses. This approach allowed us to leverage the rich chemical information captured in the MDs and MEs while maintaining the ability of the ML/DL models to represent non-linear relationships between chemical structure and toxicity across different concentration levels.

The Random Forest model was implemented using scikit-learn (version 1.7.2) with careful hyperparameter optimization to ensure robust performance. We constructed the training dataset for the ME and FME by combining the 384-dimensional ChemBERTa embeddings, and for the MD and FMD by combining the 103 descriptors, with concentration values for each compound, resulting in a feature vector that incorporated both chemical structure information and exposure concentration. The target variable was the normalized cytotoxicity, which ranged from 0 (no cytotoxicity) to 1 (complete cytotoxicity).

To identify optimal model parameters, we performed an extensive grid search over key hyperparameters, including the number of estimators (trees) {[}100, 200, 300, 400{]}, maximum tree depth {[}None, 10, 20, 30, 40{]}, and split criterion {[}\textquotesingle squared\_error\textquotesingle, \textquotesingle absolute\_error\textquotesingle{]}. Optimal random forest parameters for the MD-based model were 200 and 10, while for the ME-based model the parameters were 100 and None for the number of estimators and maximum tree depth, respectively. Cross-validation (5-fold) was employed throughout the optimization process to prevent overfitting, with data stratified at the compound level to ensure that different concentrations of the same compound remained in the same fold. This approach was critical for evaluating the model\textquotesingle s ability to generalize entirely new chemical structures.

\section*{3. Results}

\subsection*{3.1. ChemBERTa-2 Fine-Tuning}

The fine-tuning of the ChemBERTa-2 model on the HepG2 toxicity outcomes in PubChem, was preformed to create a tailored model for the creation of toxicology domain-specific embeddings. Evaluation metrics to measure the performance of the categorical task included Accuracy, ROC-AUC, and the F1-Score, which are important for a robust evaluation of the fine-tuning process. Accuracy serves as a straightforward measure of the model's overall correctness as it reflects the ratio of true predictions to total predictions. The ROC-AUC metric accounts for the sensitivity and specificity of the classifier, which offers the ability to distinguish between toxic and non-toxic compounds over various threshold values. Lastly, the F1-Score is used for analyzing models that might suffer from class imbalances. This metric combines both precision and recall into a single metric thereby capturing the balance between false positives and false negatives. The use of all three metrics provides insight into the classifier\textquotesingle s strengths and weaknesses and ensures that one is not overly focused on a single performance metric.

\begin{longtable}[]{@{}
  >{\raggedright\arraybackslash}p{(\columnwidth - 6\tabcolsep) * \real{0.2812}}
  >{\raggedright\arraybackslash}p{(\columnwidth - 6\tabcolsep) * \real{0.2188}}
  >{\raggedright\arraybackslash}p{(\columnwidth - 6\tabcolsep) * \real{0.2500}}
  >{\raggedright\arraybackslash}p{(\columnwidth - 6\tabcolsep) * \real{0.2500}}@{}}
\toprule\noalign{}
\multicolumn{4}{@{}>{\raggedright\arraybackslash}p{(\columnwidth - 6\tabcolsep) * \real{1.0000} + 6\tabcolsep}@{}}{%
\begin{minipage}[b]{\linewidth}\raggedright
\textbf{Table 1.} Fine-Tuning of ChemBERTa-2 on PubChem Data.
\end{minipage}} \\
\midrule\noalign{}
\endhead
\bottomrule\noalign{}
\endlastfoot
\textbf{Model} & \textbf{Accuracy} & \textbf{ROC-AUC} & \textbf{F1-Score} \\
ChemBERTa Classifier & 0.89 & 0.70 & 0.41 \\
\end{longtable}

The results of the fine-tuning process (Table 1) indicate promising metrics, achieving an accuracy of 0.89, ROC-AUC value of 0.70, and an F1-Score of 0.41.. In addition to Accuracy, ROC-AUC, and F1-score, a confusion matrix was created (Figure 1) to provide a detailed breakdown of the predictions relative to the ground truth. Inspection of the confusion matrix indicates the model has strong performance in predicting non-toxic molecules correctly (0.93) and a low probability of predicting them incorrectly (0.07). Where the model struggles in predicting toxic molecules (0.47) correctly. These limitations of the model could be related to the high-throughput nature of the HepG2 PubChem data. Inspection of the PubChem dose-response curves reveals a significant amount of noise in the recorded data likely due to the relatively small volume in the 1536-well assay plates. This noise impacts the following categorical characterization of the curve data, which impacts the fine-tuning process.

\subsection*{3.2 Model Validation Results}

We trained three Random Forest (RF) regression models to predict dose-response curves for chemical compounds and mixtures. These models were built using different molecular representations: a molecular descriptor-based RF model (MD RF), a pre-trained molecular embedding RF model (pre-trained ME RF), and a fine-tuned molecular embedding RF model (fine-tuned ME RF). The validation of the three Random Forest (RF) models was conducted using Kfold (5-fold) cross-validation on the training data set (n=554).

\begin{figure}[htbp]
\centering
\includegraphics[width=0.62\textwidth,height=0.60\textheight,keepaspectratio]{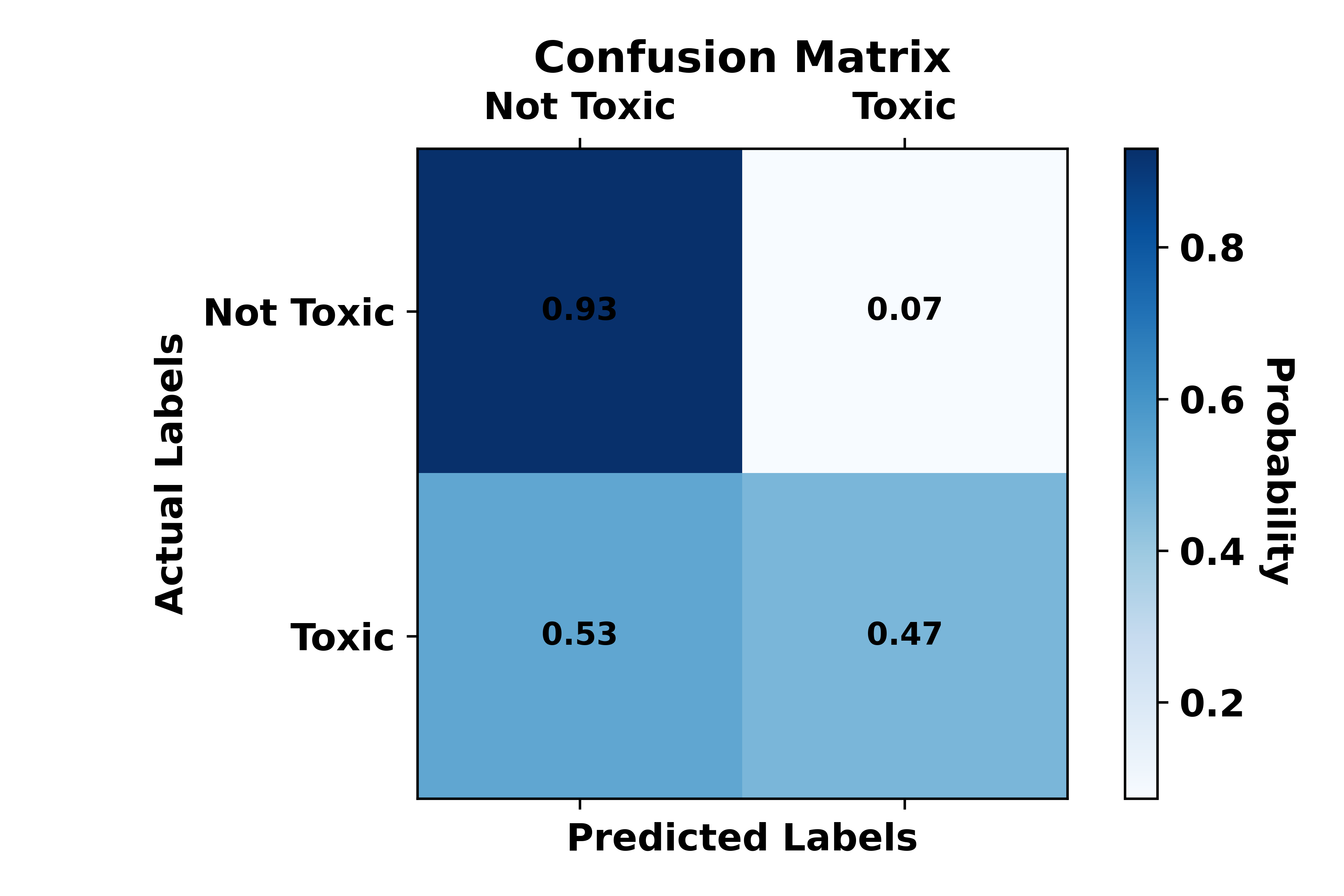}
\caption*{\textbf{Figure 1.} Confusion matrix for the ChemBERTa-based classification model predicting HepG2 toxicity outcome.}
\end{figure}
\FloatBarrier

To quantify model accuracy, three widely used metrics were calculated (Table 2): R-squared (R\textsuperscript{2}), Mean Absolute Error (MAE), and Root Mean Squared Error (RMSE). While R\textsuperscript{2} emphasizes the global fit of the model, MAE provides a more intuitive measure of average predictive error, and RMSE prioritizes sensitivity to larger deviations. Each metric holds strengths and weaknesses, underscoring the importance of using all three in the validation of a model and the comparison of model performance between models. Evaluation of the validation metrics for the three models indicate that both the MD based and fine-tuned EM models yielded strong results with an R\textsuperscript{2} of 0.89 \ensuremath{\pm} 0.05, while the pre-trained ME RF model had similar performance with an R\textsuperscript{2} of 0.90 \ensuremath{\pm} 0.04. These values indicate that all models are effectively explaining a high proportion of the variance in the target variable. The MAE values were similarly low with values of 0.066 \ensuremath{\pm} 0.02, 0.065 \ensuremath{\pm} 0.02, and 0.068 \ensuremath{\pm} 0.02 for the MD RF model, pre-trained ME RF, and the fine-tuned ME RF models, respectively. These results indicate comparable predictive accuracy in terms of average errors. Likewise, RMSE values were consistent at 0.14 \ensuremath{\pm} 0.03 for both MD RF and fine-tuned ME RF models, and 0.13 \ensuremath{\pm} 0.03 for the pre-trained ME RF model, which highlights the model's ability to control larger prediction errors.

\begin{longtable}[]{@{}
  >{\raggedright\arraybackslash}p{(\columnwidth - 6\tabcolsep) * \real{0.2500}}
  >{\raggedright\arraybackslash}p{(\columnwidth - 6\tabcolsep) * \real{0.2500}}
  >{\raggedright\arraybackslash}p{(\columnwidth - 6\tabcolsep) * \real{0.2500}}
  >{\raggedright\arraybackslash}p{(\columnwidth - 6\tabcolsep) * \real{0.2500}}@{}}
\toprule\noalign{}
\multicolumn{4}{@{}>{\raggedright\arraybackslash}p{(\columnwidth - 6\tabcolsep) * \real{1.0000} + 6\tabcolsep}@{}}{%
\begin{minipage}[b]{\linewidth}\raggedright
\textbf{Table 2.} Kfold Cross Validation Results on Train Data Set (n=554).
\end{minipage}} \\
\midrule\noalign{}
\endhead
\bottomrule\noalign{}
\endlastfoot
\textbf{Model} & \textbf{R\textsuperscript{2}} & \textbf{MAE\textsuperscript{1}} & \textbf{RMSE\textsuperscript{1}} \\
MD RF & 0.89 \ensuremath{\pm} 0.05 & 0.066 \ensuremath{\pm} 0.02 & 0.14 \ensuremath{\pm} 0.03 \\
Pre-trained ME RF & \textbf{0.90 \ensuremath{\pm} 0.04} & \textbf{0.065 \ensuremath{\pm} 0.02} & \textbf{0.13 \ensuremath{\pm} 0.03} \\
Fine-tuned ME RF & 0.89 \ensuremath{\pm} 0.05 & 0.068 \ensuremath{\pm} 0.02 & 0.14 \ensuremath{\pm} 0.03 \\
\end{longtable}

\textsuperscript{1} Units are Cytotoxicity.

\subsection*{3.3 Dose Response Curve Preditions}

The three model's performance on the test set of data (20\% hold out) was evaluated for R\textsuperscript{2}, MAE, and RMSE and is found in Table 3. The results indicate that the use of MD, pre-trained ME, and fine-tuned ME as input features to the RF model framework yield highly accurate models with little distinction between the treatment of the input features. Plotting the predicted values against the ground truth (Figure 2) indicates that all models accurately predict points at the extremes. Where the models appear to incur the most error is in the central region, which is not surprising given this is the region with largest change

\begin{longtable}[]{@{}
  >{\raggedright\arraybackslash}p{(\columnwidth - 6\tabcolsep) * \real{0.2500}}
  >{\raggedright\arraybackslash}p{(\columnwidth - 6\tabcolsep) * \real{0.2500}}
  >{\raggedright\arraybackslash}p{(\columnwidth - 6\tabcolsep) * \real{0.2500}}
  >{\raggedright\arraybackslash}p{(\columnwidth - 6\tabcolsep) * \real{0.2500}}@{}}
\toprule\noalign{}
\multicolumn{4}{@{}>{\raggedright\arraybackslash}p{(\columnwidth - 6\tabcolsep) * \real{1.0000} + 6\tabcolsep}@{}}{%
\begin{minipage}[b]{\linewidth}\raggedright
\textbf{Table 3.} Random Forest Model Metrics on Test Data Set (n=139)
\end{minipage}} \\
\midrule\noalign{}
\endhead
\bottomrule\noalign{}
\endlastfoot
\textbf{Model} & \textbf{R\textsuperscript{2}} & \textbf{MAE\textsuperscript{1}} & \textbf{RMSE\textsuperscript{1}} \\
MD RF & 0.90 & 0.056 & 0.12 \\
Pre-trained ME RF & \textbf{0.92} & \textbf{0.046} & \textbf{0.11} \\
Fine-tuned ME RF & 0.91 & 0.052 & 0.12 \\
\multicolumn{4}{@{}>{\raggedright\arraybackslash}p{(\columnwidth - 6\tabcolsep) * \real{1.0000} + 6\tabcolsep}@{}}{%
\textsuperscript{1} Units are Cytotoxicity.} \\
\end{longtable}

in slope. Evaluating the actual dose response curves for individual compounds (Figure 3) and for mixtures (Figure 4) provides additional insight into the strength and weakness of the models. A detailed analysis of each curve can be found in the Supplemental information (Supplemental Section S.3, S.4, and S.5). Figure 3 indicates that the pre-trained ME RF and fine-tuned ME RF appear to have small improvements in the predictive accuracy in the intermediate concentration ranges. It was found that the use of pre-trained ME and fine-tuned ME yields a marginally better fit for metals and OPFRS while there was no observable improvement for PFOAs and phenols. In the case of the mixture predictions there appears to be an improved ability of the pre-trained ME RF model to predict the maximal cytotoxicity with respect to the fine-tuned ME RF model and to a similar accuracy as the MD RF model.

\begin{figure}[htbp]
\centering
\includegraphics[width=0.52\textwidth,height=0.74\textheight,keepaspectratio]{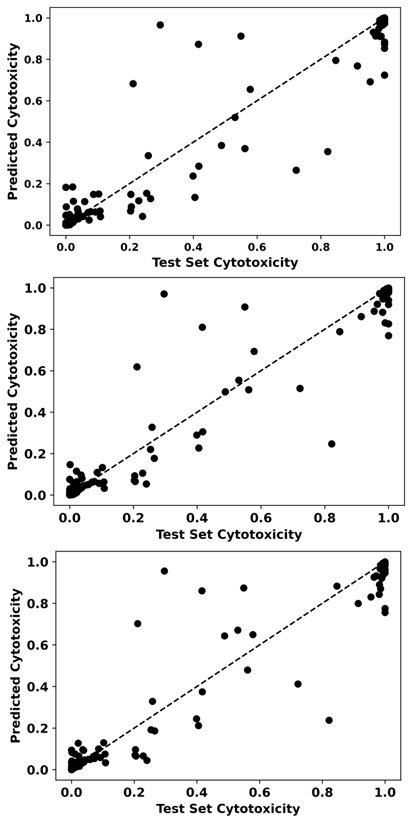}
\caption*{\textbf{Figure 2.} Test set ground truth against (Top) MD RF model (Middle) pre-trained ME RF model, and (Bottom) fine-tuned ME RF model.}
\end{figure}
\FloatBarrier

\begin{figure}[htbp]
\centering
\includegraphics[width=0.52\textwidth,height=0.74\textheight,keepaspectratio]{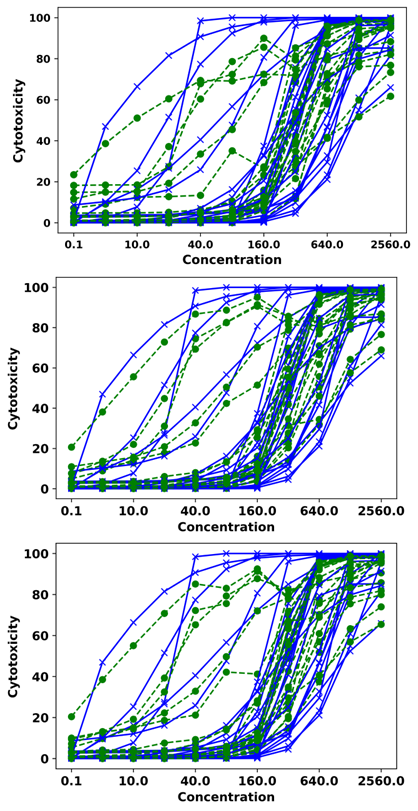}
\caption*{\textbf{Figure 3.} Individual compound dose response curve for (Top) MD RF, (Middle) pre-trained ME RF, and (Bottom) fine-tuned ME RF where blue ($\chi$) is ground truth and the green ($\bullet$) are predicted values.}
\end{figure}
\FloatBarrier

\section*{4. Discussion}

Our comparative analysis of different molecular representation approaches for liver toxicity prediction revealed similar performance across methods, with pre-trained ChemBERTa embeddings combined with Random Forest regression (R\textsuperscript{2} = 0.92) performing slightly better than fine-tuned ChemBERTa embeddings (R\textsuperscript{2} = 0.91) and traditional RCDK descriptors (R\textsuperscript{2} = 0.90). This finding is noteworthy as it suggests that pre-trained transformer-based embeddings inherently capture chemical patterns relevant to toxicological outcomes without requiring endpoint-specific fine-tuning. The contextual nature of the transformer architecture allows ChemBERTa to learn meaningful representations of chemical substructures and their interactions within the broader molecular context, achieving comparable performance to traditional descriptor-based methods that rely on predefined chemical features.

\begin{figure}[htbp]
\centering
\includegraphics[width=0.52\textwidth,height=0.74\textheight,keepaspectratio]{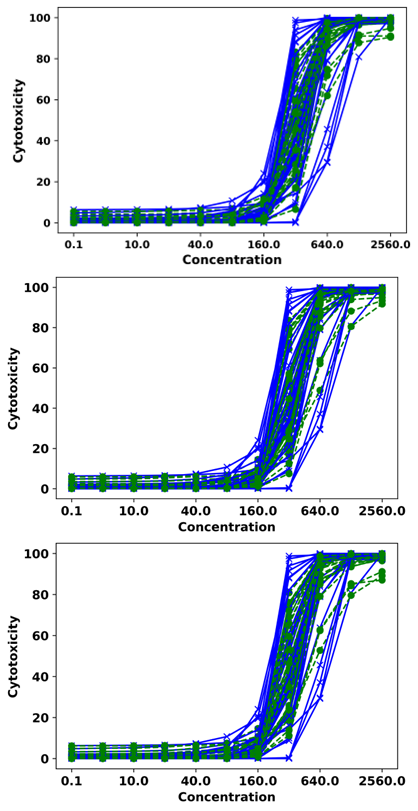}
\caption*{\textbf{Figure 4.} Mixture dose response curve for (Top) MD RF, (Middle) ME RF, and (Bottom) fine-tuned ME RF where blue (x) is ground truth and the green ($\bullet$) are predicted values.}
\end{figure}
\FloatBarrier

Interestingly, the fine-tuning of ChemBERTa on HepG2 toxicity data did not substantially improve performance beyond what was achieved with the pre-trained embeddings. This result may be attributed to limitations in the fine-tuning dataset, which was derived from high-throughput experiments using microwell-plates with relatively few cells per well. Such experimental conditions can introduce variability and noise that potentially compromised the quality of the training signal during fine-tuning. This observation underscores the importance of high-quality training data for effective model fine-tuning and suggests that in cases where endpoint-specific data may be of limited quality, leveraging pre-trained embeddings directly can be an equally effective strategy.

Perhaps the most significant innovation in this approach is the successful prediction of mixture toxicity using weighted molecular embeddings. By combining individual compound embeddings weighted by their mole fractions, we were able to generate representations of mixtures that effectively capture their toxicity profiles, as evidenced by the strong performance of this approach, a high R\textsuperscript{2} score and low prediction errors. Traditional approaches to mixture toxicity prediction often require extensive experimental data on specific mixtures or rely on simplistic concentration addition or independent action models. Our embedding-based approach offers a more flexible alternative that can potentially predict the toxicity of a combination of compounds for which individual embeddings are available, without requiring explicit mixture training data. This capability is particularly valuable for real-world risk assessment scenarios, where exposures typically involve complex mixtures rather than individual compounds.

The ability of our model to generate accurate concentration-response curves represents another significant advancement over binary classification approaches that simply categorize compounds as toxic or non-toxic. By predicting toxicity across different concentration levels, our approach provides a more comprehensive assessment of chemical hazard that aligns with the dose-dependent nature of toxicity. This capability enables the calculation of important toxicological parameters such as EC10 and EC50 values, which are essential for regulatory decision-making and risk assessment. The close agreement between predicted and experimental concentration-response curves for both individual compounds and mixtures demonstrates the robustness of our approach across different scenarios.

Despite these promising results, several limitations must be acknowledged. The current approach assumes additivity in the embedding space for mixture toxicity prediction, which may not fully capture synergistic or antagonistic effects that can occur in complex mixtures. Additionally, the training data from HepG2 cell line experiments may not fully represent the complexity of liver tissue responses \emph{in vivo}, particularly regarding metabolic activation or detoxification processes. The imbalanced nature of the fine-tuning training dataset, with significantly more non-toxic than toxic compounds, could potentially bias the model toward predicting non-toxicity. Furthermore, while the embedding space provides rich information about chemical-toxicity relationships, the interpretability of individual embedding dimensions remains challenging, limiting mechanistic insights into toxicity pathways.

Future work should focus on addressing these limitations through several approaches. Integration of metabolic activation data could improve the biological relevance of the predictions, particularly for compounds that require metabolic transformation to exert their toxic effects. Expansion to additional cell types and toxicity endpoints would broaden the applicability of the approach beyond liver toxicity. Development of more sophisticated mixture models that explicitly account for synergistic and antagonistic interactions could further improve mixture toxicity predictions.

\section*{5. Conclusions}

This study introduces a transformative approach to predictive toxicology by leveraging ChemBERTa-based molecular representations for liver toxicity prediction. The findings reveal that pre-trained and fine-tuned ChemBERTa embeddings paired with Random Forest regression performs at the same level as a high performing traditional descriptor-based methods. This result demonstrates the inherent capacity of transformer-based architectures to capture chemical features relevant to toxicological outcomes without endpoint-specific fine-tuning. Furthermore, the use of weighted molecular embeddings for mixture toxicity prediction represents a viable methodology for the characterization of complex mixture toxicity profiles. This approach addresses key challenges in real-world risk assessment and regulatory applications where exposure to chemical mixtures is more common than individual compounds. Additionally, the ability to predict complete concentration-response curves adds significant benefits such as allowing dose-dependent hazard evaluation and the calculation of critical toxicological parameters like EC10 and EC50 values.

\section*{Abbreviations}

The following abbreviations are used in this manuscript:

\begin{longtable}[]{@{}
  >{\raggedright\arraybackslash}p{(\columnwidth - 2\tabcolsep) * \real{0.1191}}
  >{\raggedright\arraybackslash}p{(\columnwidth - 2\tabcolsep) * \real{0.8809}}@{}}
\toprule\noalign{}
\begin{minipage}[b]{\linewidth}\raggedright
AI
\end{minipage} & \begin{minipage}[b]{\linewidth}\raggedright
Artificial Intelligence
\end{minipage} \\
\midrule\noalign{}
\endhead
\bottomrule\noalign{}
\endlastfoot
CA & Concentration Addition \\
DL & Deep Learning \\
FMD & Formula Molecular Descriptors \\
FME & Formula Molecular Embeddings \\
IA & Independent Action \\
LLM & Large Language Model \\
MoA & Mode of Action \\
MD & Molecular Descriptors \\
ME & Molecular Embeddings \\
ML & Machine Learning \\
OPFRS & Organophosphate Flame Retardants \\
PFAS & Polyfluoroalkyl Compounds \\
RF & Random Forest \\
QSAR & Quantitative Structure Activity Relationship \\
RCDK & R's Chemistry Development Kit \\
ROC-AUC & Receiver Operating Characteristic Area under the Curve \\
SMILES & Simplified Molecular Input Line Entry System \\
2,4-DCP & 2,4-dichlorophenol \\
2,5-DCP & 2,5-dichlorophenol \\
BP-3 & Benzophenone-3 \\
bPB & Butyl paraben \\
Cd & Cadmium chloride hydrate \\
Co & Cobalt chloride \\
Cu & Cupric sulfate \\
EHDPHP & 2-Ethylhexyl diphenyl phosphate \\
Hg & Methylmercury chloride \\
Ni & Nickel dichloride \\
Pb & Lead chloride \\
PFHxS & Perfluorohexanesulfonic acid \\
PFNA & Perfluorononanoic acid \\
PFOA & Perfluorooctanoic acid \\
pPB & Propyl paraben \\
Sb & Antimony(III) chloride \\
Se & Sodium selenite \\
TBOEP & Tris(2-butoxyethyl) phosphate \\
TCPP & Tris(1-chloro-2-propyl)phosphate \\
TCS & Triclosan \\
TDCPP & Tris(1,3-dichloropropyl) phosphate \\
TEHP & Tri (2-ethylhexyl)phosphate \\
TPhP & Triphenyl phosphate \\
Zn & Zinc sulfate heptahydrate \\
\end{longtable}

Supplementary Materials: The following supporting information can be downloaded at: \url{https://www.mdpi.com/article/doi/s1}, Table S1: List of Compounds, Table S2: List of Mixtures, Figure S3.1: Dose Response Curves for MD RF Model of Heavy Metals, Figure S3.2: Dose Response Curves for MD RF Model of Organophosphate Flame Retardants, Figure S3.3: Dose Response Curves for MD RF Model of Polyfluoroalkyl Compounds, Figure S3.4: Dose Response Curves for MD RF Model of Phenols, Figure S3.5: Dose Response Curves for MD RF Model of Mixtures 1 to 28, Figure S3.6: Dose Response Curves for MD RF Model of Mixtures 30 to 48, Figure S4.1: Dose Response Curves for Pre-Trained ME RF Model of Heavy Metals, Figure S4.2: Dose Response Curves for Pre-Trained ME RF Model of Organophosphate Flame Retardants, Figure S4.3: Dose Response Curves for Pre-Trained ME RF Model of Polyfluoroalkyl Compounds, Figure S4.4: Dose Response Curves for Pre-Trained ME RF Model of Phenols, Figure S4.5: Dose Response Curves for Pre-Trained ME RF Model of Mixtures 1 to 28, Figure S4.6: Dose Response Curves for Pre-Trained ME RF Model of Mixtures 30 to 48, Figure S5.1: Dose Response Curves for Fine-Tuned ME RF Model of Heavy Metals, Figure S5.2: Dose Response Curves for Fine-Tuned ME RF Model of Organophosphate Flame Retardants, Figure S5.3: Dose Response Curves for Fine-Tuned ME RF Model of Polyfluoro-alkyl Compounds, Figure S5.4: Dose Response Curves for Fine-Tuned ME RF Model of Phenols, Figure S5.5: Dose Response Curves for Fine-Tuned ME RF Model of Mixtures 1 to 28, Figure S5.6: Dose Response Curves for Fine-Tuned ME RF Model of Mixtures 30 to 48.

Author Contributions: Conceptualization, C.M.M., S.P.D, S.D. and A.S.; methodology, C.M.M., S.P.D, S.D., and A.S.; software, C.M.M and A.S.; validation, C.M.M., J.Z, S.S and A.S.; investigation, C.M.M., J.Z., S.D., S.P.D., Z.R.S., B.H., S.S., and A.S.; data curation, C.M.M., J.Z. and A.S.; writing---original draft preparation, C.M.M. and A.S.; writing---review and editing, C.M.M., J.Z., S.P.D., S.D., Z.R.S, B.H., S.S. and A.S; All authors have read and agreed to the published version of the manuscript.

Funding: The research described in this paper was conducted under the Laboratory Directed Re-search and Development Program at Pacific Northwest National Laboratory, a multiprogram national laboratory operated by Battelle for the U.S. Department of Energy.

Data Availability Statement: All data and models available at: \url{https://github.com/pnnl/Finch}

Acknowledgments: A portion of the research was performed using resources available through Research Computing at Pacific Northwest National Laboratory (PNNL). PNNL is operated by Battelle for the U.S. Department of Energy under Contract DE-AC05-76RL01830. During the preparation of this manuscript, the authors used OpenAI's gpt 4.o for the purposes of assisting in manuscript writing and creating readme documents for the code. The authors have reviewed and edited the output and take full responsibility for the content of this publication.

Conflicts of Interest: The authors declare no conflicts of interest.

\section*{References}

1. Sarigiannis, D.A. and U. Hansen, \emph{Considering the cumulative risk of mixtures of chemicals -- A challenge for policy makers.} Environmental Health, 2012. \textbf{11}(1): p. S18.

2. Shaw, I.C., \emph{Chemical residues, food additives and natural toxicants in food -- the cocktail effect.} International Journal of Food Science and Technology, 2014. \textbf{49}(10): p. 2149-2157.

3. Wang, N., et al., \emph{Prediction of the joint action of binary mixtures based on characteristic parameter k\ensuremath{\cdot}ECx from concentration-response curves.} Ecotoxicology and Environmental Safety, 2021. \textbf{215}: p. 112155.

4. Cedergreen, N., \emph{Quantifying synergy: a systematic review of mixture toxicity studies within environmental toxicology.} PLoS One, 2014. \textbf{9}(5): p. e96580.

5. Elcombe, C.S., E.N. P., and M. Bellingham, \emph{Critical review and analysis of literature on low dose exposure to chemical mixtures in mammalian in~vivo systems.} Critical Reviews in Toxicology, 2022. \textbf{52}(3): p. 221-238.

6. Tropsha, A., et al., \emph{Integrating QSAR modelling and deep learning in drug discovery: the emergence of deep QSAR.} Nature Reviews Drug Discovery, 2024. \textbf{23}(2): p. 141-155.

7. Sabando, M.V., et al., \emph{Using molecular embeddings in QSAR modeling: does it make a difference?} Briefings in Bioinformatics, 2021. \textbf{23}(1).

8. Colby, S.M., et al., \emph{Deep Learning to Generate in Silico Chemical Property Libraries and Candidate Molecules for Small Molecule Identification in Complex Samples.} Analytical Chemistry, 2020. \textbf{92}(2): p. 1720-1729.

9. Chen, J.-H. and Y.J. Tseng, \emph{A general optimization protocol for molecular property prediction using a deep learning network.} Briefings in Bioinformatics, 2021. \textbf{23}(1).

10. Wang, T., et al., \emph{Prediction of the Toxicity of Binary Mixtures by QSAR Approach Using the Hypothetical Descriptors.} International Journal of Molecular Sciences, 2018. \textbf{19}(11): p. 3423.

11. Zhang, F., et al., \emph{Machine learning-driven QSAR models for predicting the mixture toxicity of nanoparticles.} Environ Int, 2023. \textbf{177}: p. 108025.

12. Kim, J., M. Seo, and M. Na, \emph{MRA Toolbox v. 1.0: a web-based toolbox for predicting mixture toxicity of chemical substances in chemical products.} Scientific Reports, 2022. \textbf{12}(1): p. 8880.

13. Chatterjee, M. and K. Roy, \emph{Predictive binary mixture toxicity modeling of fluoroquinolones (FQs) and the projection of toxicity of hypothetical binary FQ mixtures: a combination of 2D-QSAR and machine-learning approaches.} Environmental Science: Processes \& Impacts, 2024. \textbf{26}(1): p. 105-118.

14. Chatterjee, M. and K. Roy, \emph{Prediction of aquatic toxicity of chemical mixtures by the QSAR approach using 2D structural descriptors.} Journal of Hazardous Materials, 2021. \textbf{408}: p. 124936.

15. Abbod, M. and A. Mohammad, \emph{Combined interaction of fungicides binary mixtures: experimental study and machine learning-driven QSAR modeling.} Scientific Reports, 2024. \textbf{14}(1): p. 12700.

16. Kim, S., et al., \emph{In Vitro Toxicity Screening of Fifty Complex Mixtures in HepG2 Cells.} Toxics, 2024. \textbf{12}(126): p. 1-12.

17. Ge, H., et al., \emph{Integrative Assessment of Mixture Toxicity of Three Ionic Liquids on Acetylcholinesterase Using a Progressive Approach from 1D Point, 2D Curve, to 3D Surface.} Int J Mol Sci., 2019. \textbf{20}(21): p. 5330.

18. Singh, S.P., \emph{Transfer of learning by composing solutions of elemental sequential tasks.} Machine Learning, 1992. \textbf{8}(3): p. 323-339.

19. Simões, R.S., et al., \emph{Transfer and Multi-task Learning in QSAR Modeling: Advances and Challenges.} Front Pharmacol, 2018. \textbf{9}: p. 74.

20. Pratt, L. and B. Jennings, \emph{A Survey of Transfer Between Connectionist Networks.} Connection Science, 1996. \textbf{8}(2): p. 163-184.

21. Steinbeck, C., et al., \emph{The Chemistry Development Kit (CDK): An Open-Source Java Library for Chemo and Bioinformatics.} J. Chem. Inf. Comput. Sci., 2003. \textbf{43}: p. 493-500.

\end{document}


{\large\bfseries Supplemental}\par
\vspace{0.5em}
{\LARGE\bfseries Finch: Toxicity Dose Response Curve Prediction of Chemical Compounds and Mixtures}\par
\vspace{0.7em}
Abdullah Shouaib\textsuperscript{1}, John Zapanta\textsuperscript{1}, Sean P. Davern\textsuperscript{1}, Samuel Dixon\textsuperscript{1}, Zachary Stromberg\textsuperscript{1}, Becky Hess\textsuperscript{1}, Sydney Schwartz\textsuperscript{1}, and C Mark Maupin\textsuperscript{1,*}\par
\vspace{0.5em}
\textsuperscript{1}Pacific Northwest National Laboratory, Richland, Washington 99354, USA; \href{mailto:Abdullah.shouaib@pnnl.gov}{\nolinkurl{Abdullah.shouaib@pnnl.gov}} (A.S.); \href{mailto:john.zapanta@pnnl.gov}{\nolinkurl{john.zapanta@pnnl.gov}} (J.Z.); \href{mailto:sean.davern@pnnl.gov}{\nolinkurl{sean.davern@pnnl.gov}} (S.P.D.); \href{mailto:Samuel.dixon@pnnl.gov}{\nolinkurl{Samuel.dixon@pnnl.gov}} (S.D.); \href{mailto:Zachary.stromberg@pnnl.gov}{\nolinkurl{Zachary.stromberg@pnnl.gov}} (Z.R.S.); \href{mailto:Becky.Hess@pnnl.gov}{\nolinkurl{Becky.Hess@pnnl.gov}} (B.H.); \href{mailto:schwartz.sydneyc@gmail.com}{\nolinkurl{schwartz.sydneyc@gmail.com}} (S.S.); \href{mailto:mark.maupin@pnnl.gov}{\nolinkurl{mark.maupin@pnnl.gov}} (C.M.M)\par
\textsuperscript{*}Correspondence: \href{mailto:mark.maupin@pnnl.gov}{\nolinkurl{mark.maupin@pnnl.gov}}; Tel.: +1 509-375-3738

\section*{Supplemental}
\subsection*{S.1 Data Set Composition}
\begin{longtable}{>{\RaggedRight\arraybackslash}p{0.50\textwidth} p{0.18\textwidth} p{0.22\textwidth}}
\caption*{\textbf{Table S1. List of Compounds}}\\
\toprule
\textbf{Chemical Name} & \textbf{Abbreviation} & \textbf{Group} \\
\midrule
\endfirsthead
\toprule
\textbf{Chemical Name} & \textbf{Abbreviation} & \textbf{Group} \\
\midrule
\endhead
\bottomrule
\endfoot
Cadmium chloride hydrate & Cd & Heavy metal \\
Cobalt chloride & Co & Heavy metal \\
Cupric sulfate & Cu & Heavy metal \\
Methylmercury chloride & Hg & Heavy metal \\
Nickel dichloride & Ni & Heavy metal \\
Lead chloride & Pb & Heavy metal \\
Antimony(III) chloride & Sb & Heavy metal \\
Sodium selenite & Se & Heavy metal \\
Zinc sulfate heptahydrate & Zn & Heavy metal \\
2-Ethylhexyl diphenyl phosphate & EHDPHP & OPFRs \\
Tris(2-butoxyethyl) phosphate & TBOEP & OPFRs \\
Tris(1-chloro-2-propyl)phosphate & TCPP & OPFRs \\
Tris(1,3-dichloropropyl) phosphate & TDCPP & OPFRs \\
Tri (2-ethylhexyl)phosphate & TEHP & OPFRs \\
Triphenyl phosphate & TPhP & OPFRs \\
Perfluorohexanesulfonic acid & PFHxS & PFASs \\
Perfluorononanoic acid & PFNA & PFASs \\
Perfluorooctanoic acid & PFOA & PFASs \\
2,4-dichlorophenol & 2,4-DCP & Phenols \\
2,5-dichlorophenol & 2,5-DCP & Phenols \\
Benzophenone-3 & BP-3 & Phenols \\
Propyl paraben & pPB & Phenols \\
Triclosan & TCS & Phenols \\
Butyl paraben & bPB & Phenols \\
\end{longtable}

\subsection*{S.2 Mixture Composition}
\textbf{Mixture Number:} Mix1, Mix2, Mix4, Mix5, Mix6, Mix7, Mix8, Mix9, Mix11, Mix12, Mix13, Mix14, Mix17, Mix18, Mix19, Mix20, Mix21, Mix22, Mix23, Mix25, Mix28, Mix30, Mix31, Mix32, Mix33, Mix34, Mix35, Mix36, Mix37, Mix38, Mix39, Mix40, Mix41, Mix42, Mix44, Mix45, Mix46, Mix47, Mix48

\begin{landscape}
\scriptsize
\setlength{\tabcolsep}{3.2pt}
\begin{longtable}{@{}ll*{9}{c}@{}}
\caption*{\textbf{Table S2. List of Mixtures}}\\
\toprule
\multicolumn{2}{l}{\textbf{Mixtures}} & \multicolumn{9}{c}{\textbf{Chemicals and their compositions (\%) in complex mixtures}} \\
\midrule
\endfirsthead
\toprule
\multicolumn{2}{l}{\textbf{Mixtures}} & \multicolumn{9}{c}{\textbf{Chemicals and their compositions (\%) in complex mixtures}} \\
\midrule
\endhead
\bottomrule
\endfoot
Mix1 & Chemicals & PFHxS & 2,4DCP & 2,5DCP & pPB & bPB & Hg & Cu & Se & Pb \\
(n=9) & MF (\%) & 7.286 & 20.358 & 18.915 & 16.714 & 11.878 & 0.716 & 0.331 & 0.261 & 23.540 \\
Mix2 & Chemicals & PFNA & PFOA & BP-3 & TCS & pPB & Hg & Cd & Sb &  \\
(n=8) & MF (\%) & 14.304 & 17.557 & 26.524 & 0.588 & 31.488 & 1.348 & 0.098 & 8.092 &  \\
Mix4 & Chemicals & TEHP & TDCPP & EHDPHP & TPhP & TnBP & TCPP & TBOEP &  &  \\
(n=7) & MF (\%) & 26.2878 & 14.065 & 8.575 & 5.273 & 16.557 & 17.389 & 11.854 &  &  \\
Mix5 & Chemicals & 2,4DCP & 2,5DCP & pPB & bPB & BP-3 & TCS & TnBP & TPhP & TCPP \\
(n=9) & MF (\%) & 19.058 & 17.707 & 15.646 & 11.120 & 13.179 & 0.292 & 9.708 & 3.093 & 10.196 \\
Mix6 & Chemicals & PFNA & PFOA & PFHxS & EHDPHP & TBOEP & TDCPP & TEHP & TPhP &  \\
(n=8) & MF (\%) & 14.451 & 17.738 & 13.868 & 10.224 & 14.134 & 16.769 & 6.529 & 6.287 &  \\
Mix7 & Chemicals & Co & Cu & Pb & Cd & Se & Zn & Sb & Hg & Ni \\
(n=9) & MF (\%) & 22.715 & 0.446 & 31.651 & 0.070 & 0.351 & 20.240 & 5.774 & 0.962 & 17.790 \\
Mix8 & Chemicals & Pb & Cd & Zn & Hg & TBOEP & TDCPP & TEHP &  &  \\
(n=7) & MF (\%) & 39.880 & 0.088 & 25.503 & 1.212 & 12.580 & 14.926 & 5.811 &  &  \\
Mix9 & Chemicals & PFNA & PFOA & PFHxS & 2,4DCP & 2,5DCP & pPB & bPB & BP-3 & TCS \\
(n=9) & MF (\%) & 7.133 & 8.755 & 6.845 & 19.125 & 17.770 & 15.701 & 11.159 & 13.218 & 0.293 \\
Mix11 & Chemicals & PFOA & PFHxS & Pb & Hg & Zn & TDCPP & TPhP & 25DCP & bPB \\
(n=9) & MF (\%) & 9.430 & 7.373 & 23.821 & 0.724 & 15.233 & 8.916 & 3.343 & 19.140 & 12.019 \\
Mix12 & Chemicals & PFNA & PFHxS & Co & Se & Hg & TnBP & TDCPP &  &  \\
(n=7) & MF (\%) & 14.620 & 14.030 & 32.533 & 0.503 & 1.378 & 19.971 & 16.966 &  &  \\
Mix13 & Chemicals & PFHxS & BP-3 & TCS & TnBP & EHDPHP & TBOEP & Cd & Sb & Ni \\
(n=9) & MF (\%) & 11.673 & 22.543 & 0.500 & 16.616 & 8.606 & 11.898 & 0.083 & 6.881 & 21.199 \\
Mix14 & Chemicals & PFOA & 25DCP & TCS & TBOEP & TDCPP & Zn & Cu & Hg &  \\
(n=8) & MF (\%) & 15.307 & 31.067 & 0.513 & 12.197 & 14.471 & 24.726 & 0.544 & 1.175 &  \\
Mix17 & Chemicals & PFHxS & BP-3 & pPB & TnBP & TPhP & Pb & Co & Cu & Ni \\
(n=9) & MF (\%) & 6.890 & 13.306 & 15.806 & 9.808 & 3.124 & 22.263 & 15.977 & 0.313 & 12.513 \\
Mix18 & Chemicals & PFNA & PFOA & PFHxS & 2,5DCP & BP-3 & Se & Ni & Pb &  \\
(n=8) & MF (\%) & 8.059 & 9.891 & 7.733 & 20.076 & 14.934 & 0.277 & 14.043 & 24.986 &  \\
Mix19 & Chemicals & bPB & 2,5DCP & TCS & TBOEP & TCPP & TEHP & Zn & Ni & Hg \\
(n=9) & MF (\%) & 14.511 & 23.108 & 0.382 & 9.072 & 13.307 & 4.191 & 18.391 & 16.164 & 0.874 \\
Mix20 & Chemicals & PFHxS & PFNA & pPB & Zn & Ni & TCPP & TnBP &  &  \\
(n=7) & MF (\%) & 8.980 & 9.357 & 20.598 & 18.553 & 16.307 & 13.424 & 12.782 &  &  \\
Mix21 & Chemicals & PFHxS & Pb & PFNA & Co & TPhP & EHDPHP & TBOEP & TDCPP &  \\
(n=8) & MF (\%) & 9.082 & 29.343 & 9.464 & 21.059 & 4.117 & 6.696 & 9.256 & 10.982 &  \\
Mix22 & Chemicals & PFHxS & 2,4DCP & 2,5DCP & bPB & BP-3 & TPhP & EHDPHP & TBOEP &  \\
(n=8) & MF (\%) & 8.223 & 22.975 & 21.347 & 13.405 & 15.879 & 3.728 & 6.062 & 8.381 &  \\
Mix23 & Chemicals & PFHxS & 2,4DCP & 2,5DCP & pPB & TCS & TnBP & TCPP & TDCPP &  \\
(n=8) & MF (\%) & 7.779 & 21.736 & 20.196 & 17.845 & 0.333 & 11.073 & 11.630 & 9.407 &  \\
Mix25 & Chemicals & 2,5DCP & pPB & Sb & PFOA & TCS & Zn & EHDPHP & TDCPP &  \\
(n=8) & MF (\%) & 24.007 & 21.212 & 5.451 & 11.828 & 0.396 & 19.106 & 6.817 & 11.182 &  \\
Mix28 & Chemicals & 2,4DCP & 2,5DCP & Sb & PFNA & PFOA & BP-3 & Ni & TEHP &  \\
(n=8) & MF (\%) & 22.320 & 20.738 & 4.709 & 8.324 & 10.217 & 15.426 & 14.506 & 3.761 &  \\
Mix30 & Chemicals & PFHxS & PFNA & TCS & Ni & TnBP & Hg & Cu & Cd &  \\
(n=8) & MF (\%) & 18.264 & 19.033 & 0.783 & 33.167 & 25.998 & 1.794 & 0.831 & 0.131 &  \\
Mix31 & Chemicals & PFHxS & Sb & TCS & Co & Ni & TBOEP & Cu & Cd &  \\
(n=8) & MF (\%) & 14.622 & 8.619 & 0.627 & 33.906 & 26.553 & 14.903 & 0.665 & 0.105 &  \\
Mix32 & Chemicals & Sb & Co & TnBP & TPhP & TCPP & TBOEP & TDCPP & Cd &  \\
(n=8) & MF (\%) & 6.922 & 27.230 & 16.716 & 5.324 & 17.556 & 11.969 & 14.200 & 0.084 &  \\
Mix33 & Chemicals & PFHxS & Sb & PFNA & Zn & TnBP & TDCPP & TEHP & Hg &  \\
(n=8) & MF (\%) & 10.321 & 6.084 & 10.756 & 21.326 & 14.692 & 12.481 & 23.326 & 1.014 &  \\
Mix34 & Chemicals & bPB & TCS & TnBP & TPhP & EHDPHP & TDCPP & Hg & Cu &  \\
(n=8) & MF (\%) & 28.905 & 0.760 & 25.238 & 8.038 & 13.071 & 21.440 & 1.741 & 0.806 &  \\
Mix35 & Chemicals & PFHxS & 2,4DCP & 2,5DCP & bPB & TCS & Co & Ni & TBOEP &  \\
(n=8) & MF (\%) & 7.566 & 21.140 & 19.642 & 12.334 & 0.324 & 17.544 & 13.739 & 7.711 &  \\
Mix36 & Chemicals & PFHxS & 2,4DCP & Sb & TCS & Zn & Ni & TCPP & EHDPHP &  \\
(n=8) & MF (\%) & 9.487 & 26.508 & 5.592 & 0.407 & 19.602 & 17.228 & 14.183 & 6.994 &  \\
Mix37 & Chemicals & bPB & Sb & PFNA & BP-3 & TnBP & TBOEP & TDCPP & Cd &  \\
(n=8) & MF (\%) & 18.417 & 6.659 & 11.772 & 21.816 & 16.081 & 11.514 & 13.661 & 0.081 &  \\
Mix38 & Chemicals & PFHxS & 2,4DCP & bPB & TnBP & TPhP & Hg & Cu & Cd &  \\
(n=8) & MF (\%) & 8.371 & 39.681 & 23.152 & 20.215 & 6.438 & 1.395 & 0.646 & 0.102 &  \\
Mix39 & Chemicals & PFNA & PFOA & TnBP & TCPP & EHDPHP & TBOEP & TEHP & Cd &  \\
(n=8) & MF (\%) & 11.250 & 13.808 & 15.367 & 16.139 & 7.959 & 11.003 & 24.398 & 0.077 &  \\
Mix40 & Chemicals & Sb & PFOA & BP-3 & TCS & Zn & TnBP & TPhP & Hg &  \\
(n=8) & MF (\%) & 7.477 & 16.224 & 24.495 & 0.544 & 26.208 & 18.056 & 5.751 & 1.246 &  \\
Mix41 & Chemicals & PFHxS & Sb & PFOA & Zn & Ni & TnBP & TBOEP & TDCPP &  \\
(n=8) & MF (\%) & 9.613 & 5.666 & 12.296 & 19.862 & 17.457 & 13.684 & 9.798 & 11.624 &  \\
Mix42 & Chemicals & PFHxS & BP-3 & TCS & Ni & TnBP & EHDPHP & Hg & Cu &  \\
(n=8) & MF (\%) & 14.096 & 27.221 & 0.604 & 25.597 & 20.065 & 10.392 & 1.384 & 0.641 &  \\
Mix44 & Chemicals & PFHxS & SB & Zn & Ni & TPhP & TBOEP & TDCPP & Cd &  \\
(n=8) & MF (\%) & 12.254 & 7.223 & 25.319 & 22.253 & 5.556 & 12.489 & 14.818 & 0.088 &  \\
Mix45 & Chemicals & bPB & PFNA & PFOA & BP-3 & EHDPHP & TEHP & Cu & Cd &  \\
(n=8) & MF (\%) & 18.251 & 11.666 & 14.319 & 21.620 & 8.254 & 25.301 & 0.509 & 0.080 &  \\
Mix46 & Chemicals & PFHxS & Sb & Co & Zn & Ni & TCPP & EHDPHP & TDCPP &  \\
(n=8) & MF (\%) & 8.903 & 5.248 & 20.645 & 19.396 & 16.168 & 13.310 & 6.564 & 10.766 &  \\
Mix47 & Chemicals & PFHxS & 2,5DCP & Co & Zn & TPhP & TCPP & TEHP & Hg &  \\
(n=8) & MF (\%) & 8.138 & 21.128 & 18.871 & 16.815 & 3.690 & 12.167 & 18.392 & 0.799 &  \\
Mix48 & Chemicals & 2,5DCP & TCS & Pb & EHDPHP & TBOEP & Hg & Cu & Cd &  \\
(n=8) & MF (\%) & 33.381 & 0.551 & 41.544 & 9.479 & 13.105 & 1.263 & 0.585 & 0.092 &  \\
\end{longtable}
*MF: Mole fraction (moles of single substance divided by total moles of mixture solution).\par
Mixture composition taken from Table S1 of Kim, S., et al., In Vitro Toxicity Screening of Fifty Complex Mixtures in HepG2 Cells. Toxics, 2024. 12(126): p. 1-12.\par
\end{landscape}

\clearpage
\subsection*{S.3 Dose Response Curves for Molecular Descriptor Random Forest Model}
\begin{figure}[!htbp]
\centering
\includegraphics[width=0.96\textwidth,keepaspectratio]{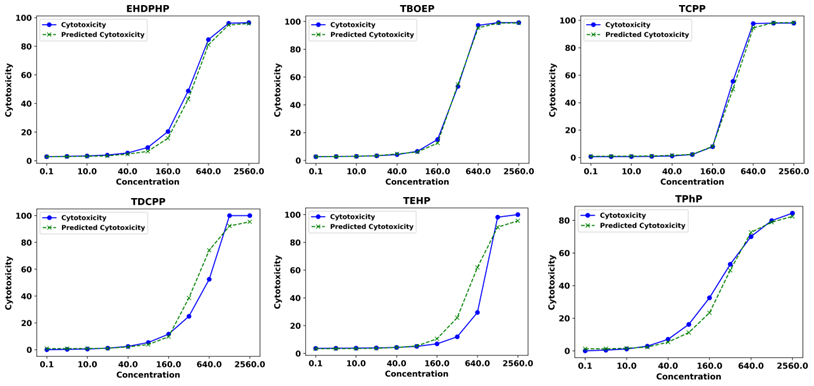}
\caption*{\textbf{Figure S3.} 2. Dose Response Curves for MD RF Model of Organophosphate Flame Retardants.}
\end{figure}

\begin{figure}[!htbp]
\centering
\includegraphics[width=0.96\textwidth,keepaspectratio]{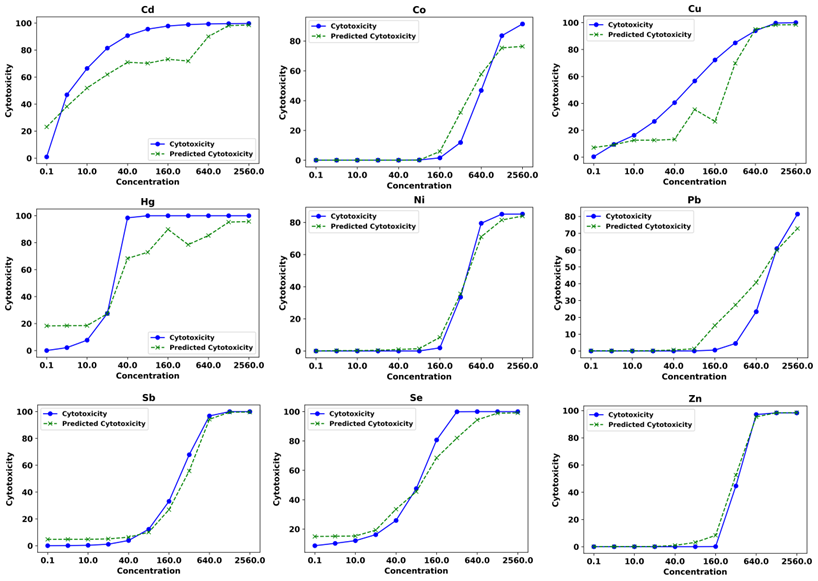}
\caption*{\textbf{Figure S3.} 1. Dose Response Curves for MD RF Model of Heavy Metals.}
\end{figure}

\begin{figure}[!htbp]
\centering
\includegraphics[width=0.96\textwidth,keepaspectratio]{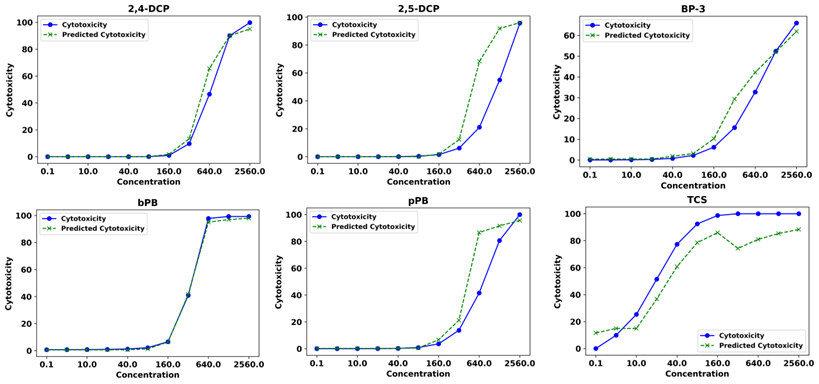}
\caption*{\textbf{Figure S3.} 4. Dose Response Curves for MD RF Model of Phenols.}
\end{figure}

\begin{figure}[!htbp]
\centering
\includegraphics[width=0.96\textwidth,keepaspectratio]{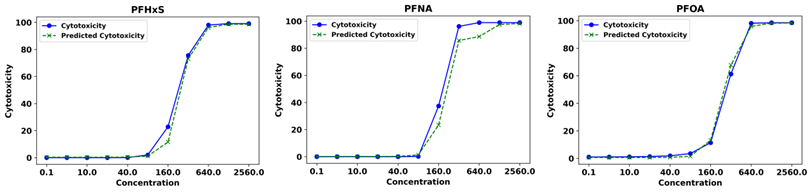}
\caption*{\textbf{Figure S3.} 3. Dose Response Curves for MD RF Model of Polyfluoroalkyl Compounds.}
\end{figure}

\begin{figure}[!htbp]
\centering
\includegraphics[height=0.78\textheight,keepaspectratio]{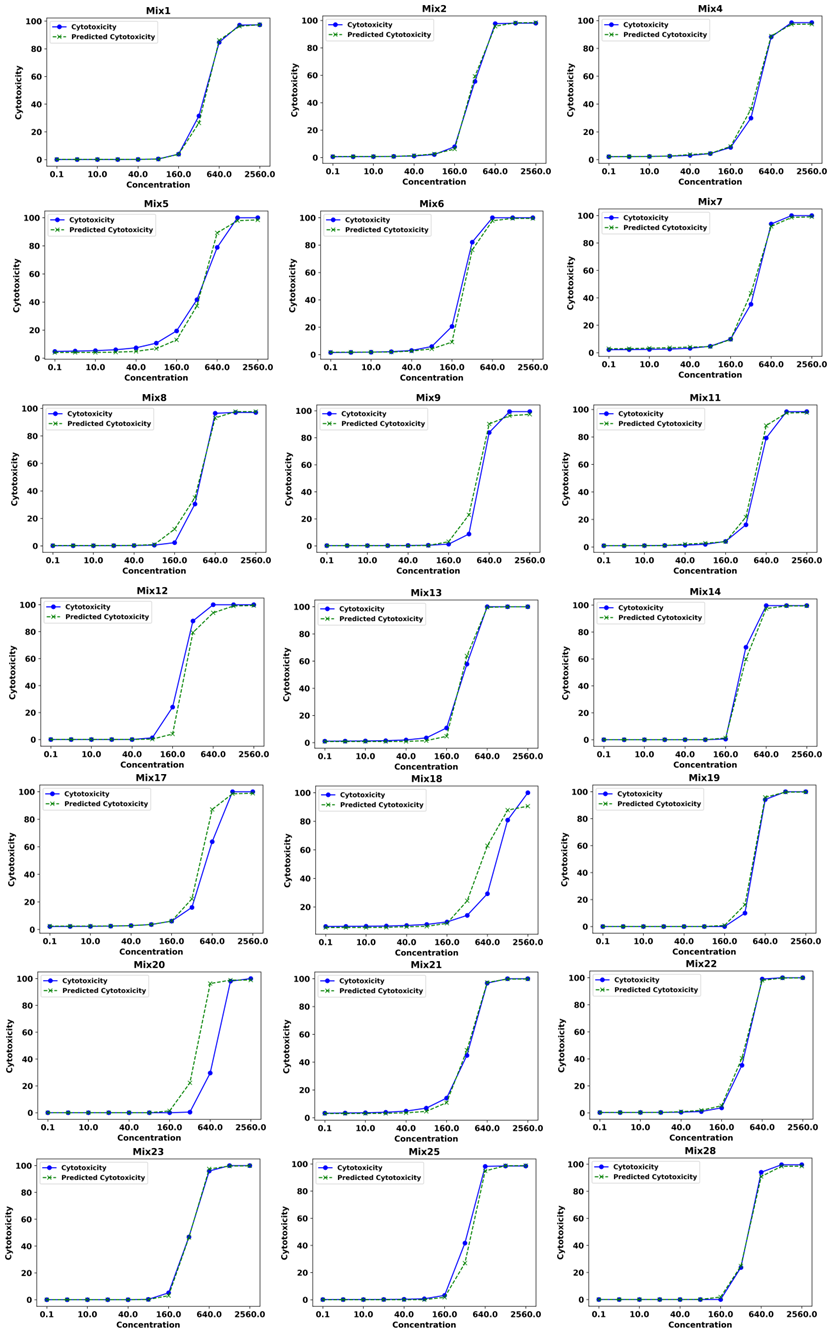}
\caption*{\textbf{Figure S3.} 5. Dose Response Curves for MD RF Model of Mixtures 1 to 28.}
\end{figure}

\begin{figure}[!htbp]
\centering
\includegraphics[height=0.78\textheight,keepaspectratio]{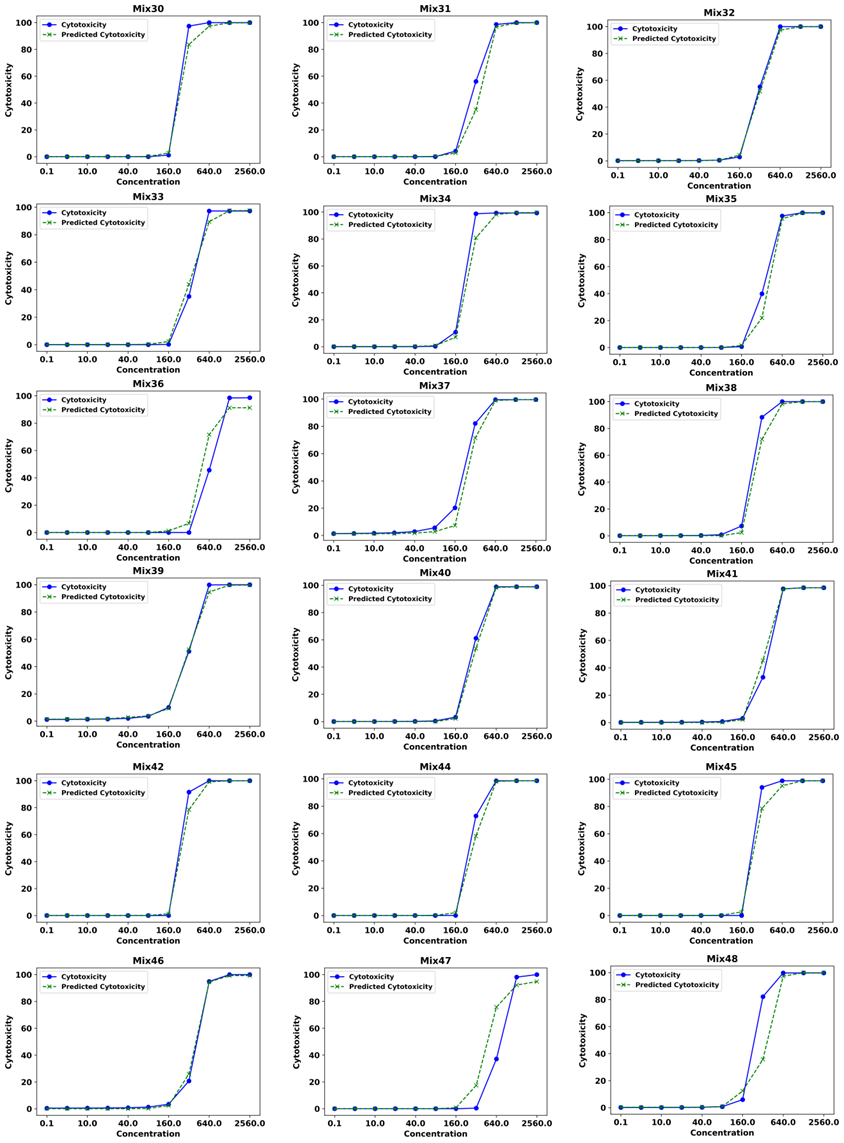}
\caption*{\textbf{Figure S3.} 6. Dose Response Curves for MD RF Model of Mixtures 30 to 48.}
\end{figure}

\clearpage
\subsection*{S.4 Dose Response Curves for Pre-Trained Molecular Embedding Random Forest Model}
\begin{figure}[!htbp]
\centering
\includegraphics[width=0.96\textwidth,keepaspectratio]{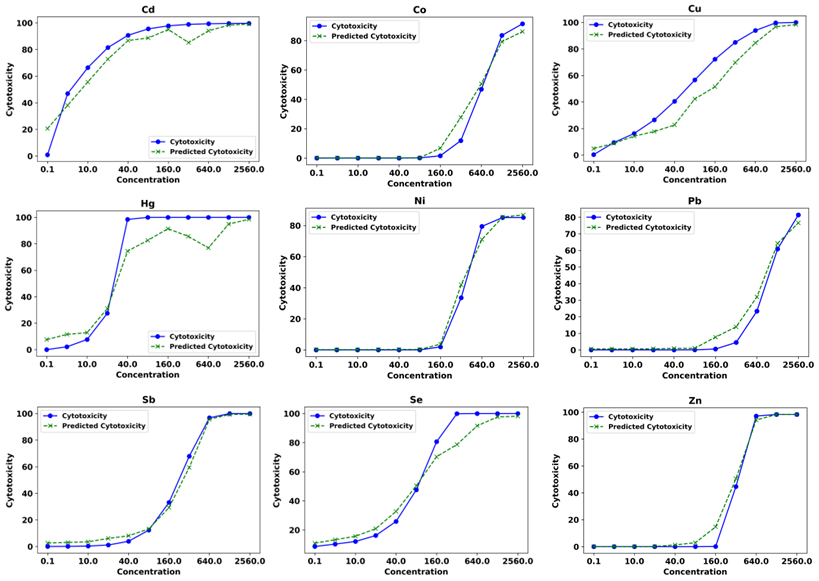}
\caption*{\textbf{Figure S4.} 1. Dose Response Curves for Pre-Trained ME RF Model of Heavy Metals.}
\end{figure}

\begin{figure}[!htbp]
\centering
\includegraphics[width=0.96\textwidth,keepaspectratio]{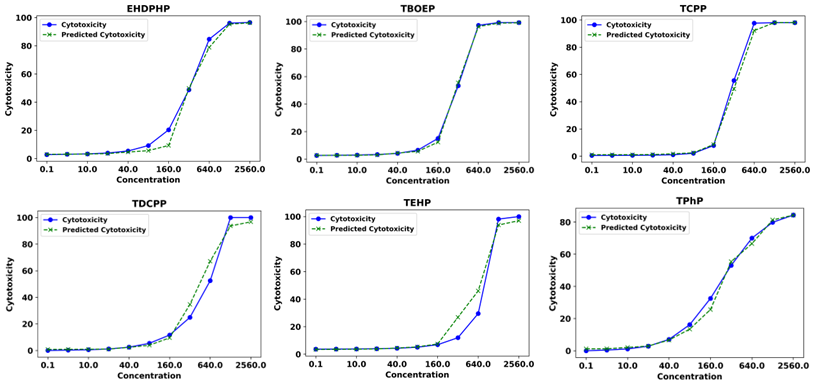}
\caption*{\textbf{Appendix Figure A4.} 2. Dose Response Curves for Pre-Trained ME RF Model of Organophosphate Flame Retardants.}
\end{figure}

\begin{figure}[!htbp]
\centering
\includegraphics[width=0.96\textwidth,keepaspectratio]{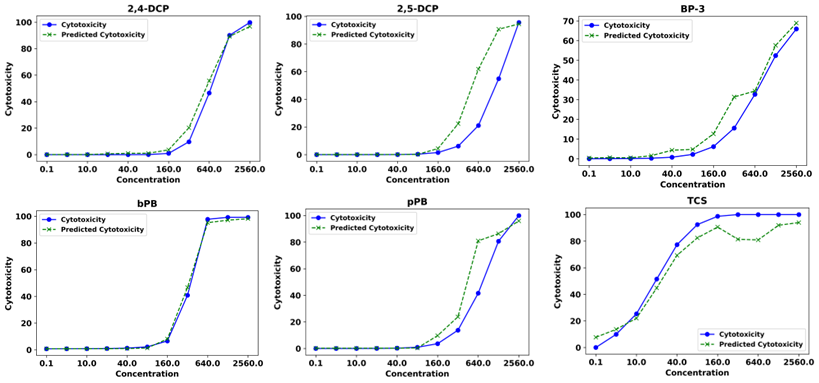}
\caption*{\textbf{Figure S4.} 4. Dose Response Curves for Pre-Trained ME RF Model of Phenols.}
\end{figure}

\begin{figure}[!htbp]
\centering
\includegraphics[width=0.96\textwidth,keepaspectratio]{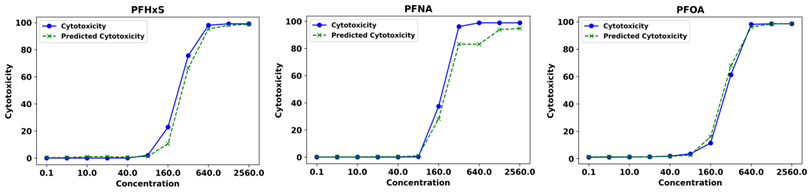}
\caption*{\textbf{Figure S4.} 3. Dose Response Curves for Pre-Trained ME RF Model of Polyfluoroalkyl Compounds.}
\end{figure}

\begin{figure}[!htbp]
\centering
\includegraphics[height=0.78\textheight,keepaspectratio]{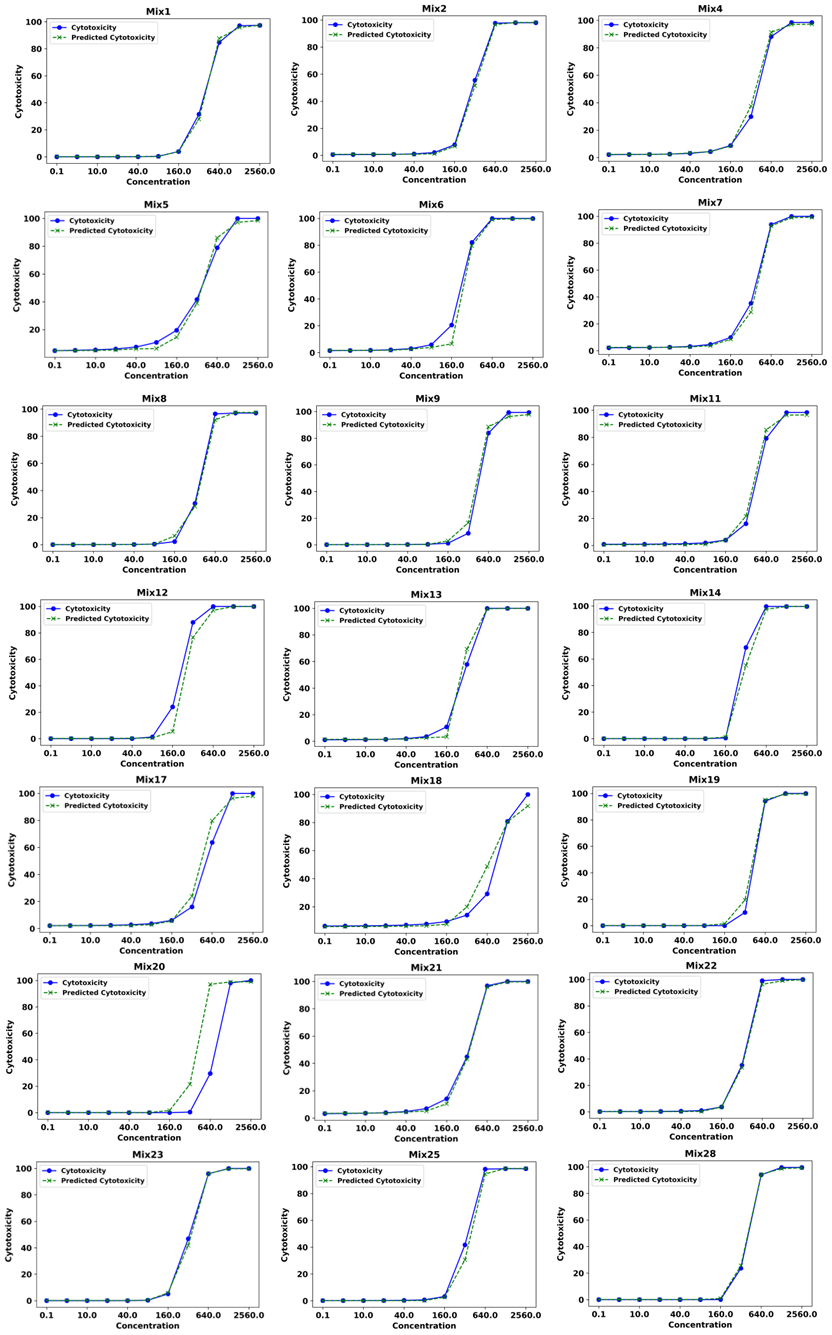}
\caption*{\textbf{Figure S4.} 5. Dose Response Curves for Pre-Trained ME RF Model of Mixtures 1 to 28.}
\end{figure}

\begin{figure}[!htbp]
\centering
\includegraphics[height=0.78\textheight,keepaspectratio]{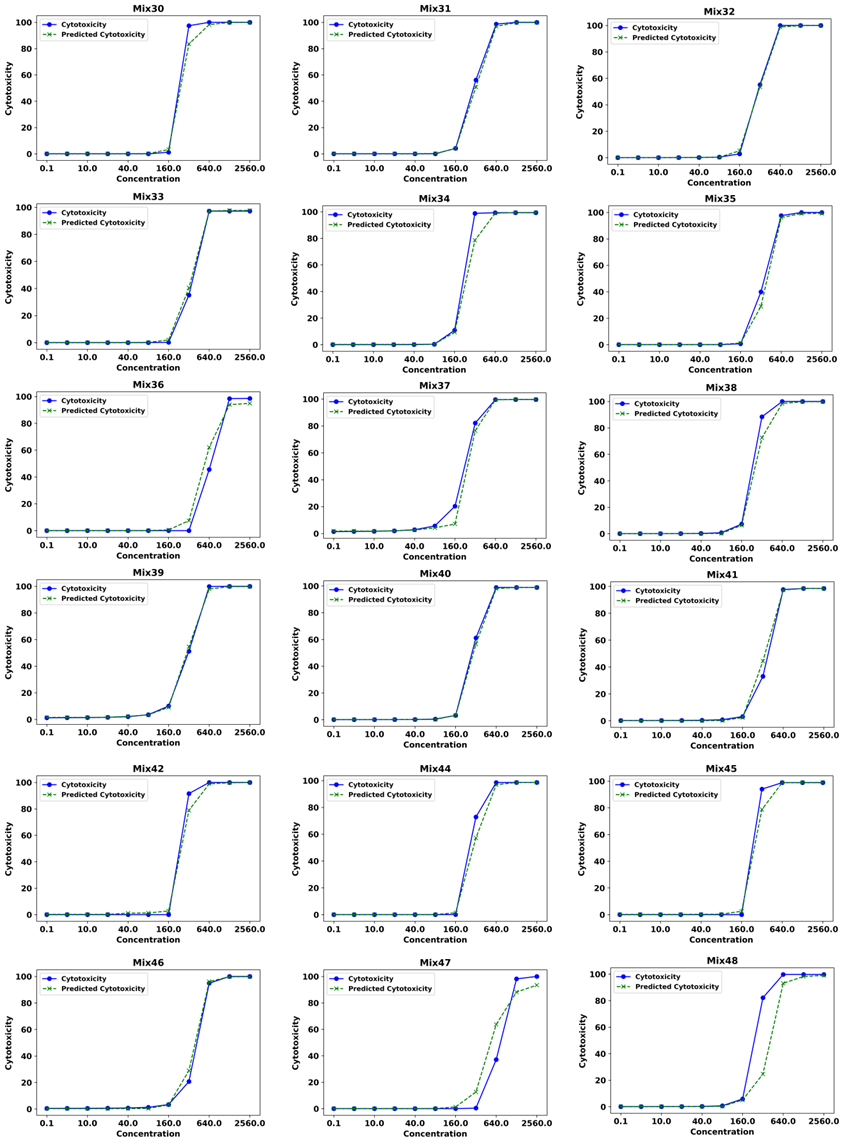}
\caption*{\textbf{Figure S4.} 6. Dose Response Curves for Pre-Trained ME RF Model of Mixtures 30 to 48.}
\end{figure}

\clearpage
\subsection*{S.5 Dose Response Curves for Fine-Tuned Molecular Embedding Random Forest Model}
\begin{figure}[!htbp]
\centering
\includegraphics[width=0.96\textwidth,keepaspectratio]{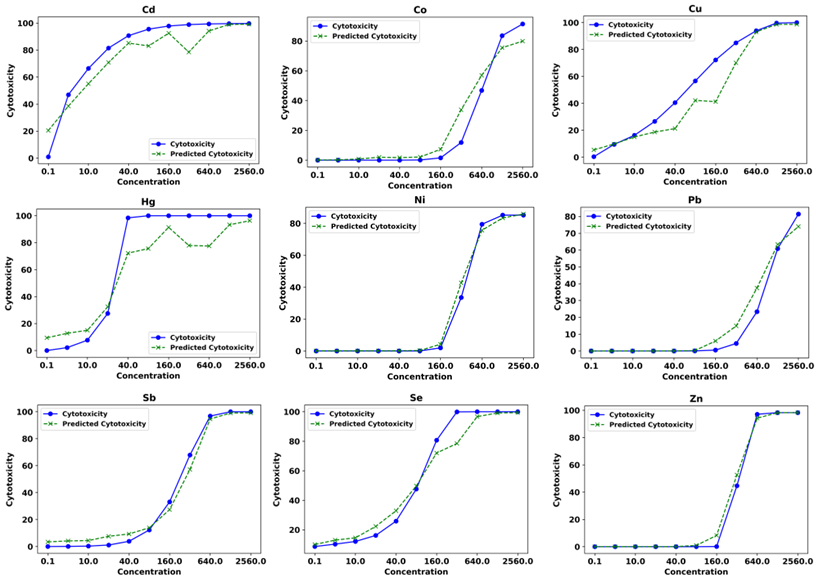}
\caption*{\textbf{Figure S5.} 1. Dose Response Curves for Fine-Tuned ME RF Model of Heavy Metals.}
\end{figure}

\begin{figure}[!htbp]
\centering
\includegraphics[width=0.96\textwidth,keepaspectratio]{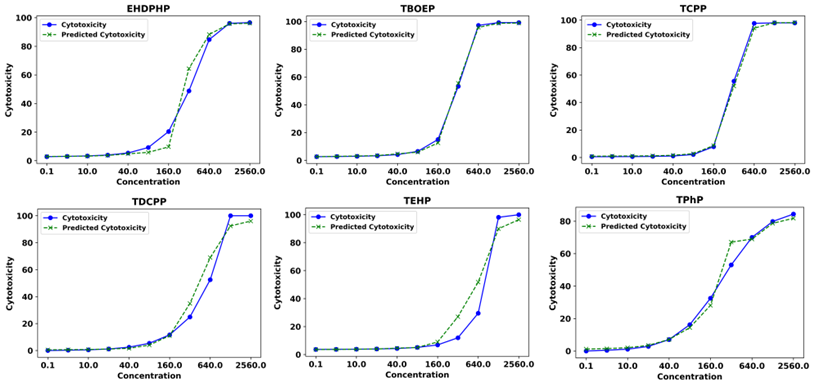}
\caption*{\textbf{Figure S5.} 2. Dose Response Curves for Fine-Tuned ME RF Model of Organophosphate Flame Retardants.}
\end{figure}

\begin{figure}[!htbp]
\centering
\includegraphics[width=0.96\textwidth,keepaspectratio]{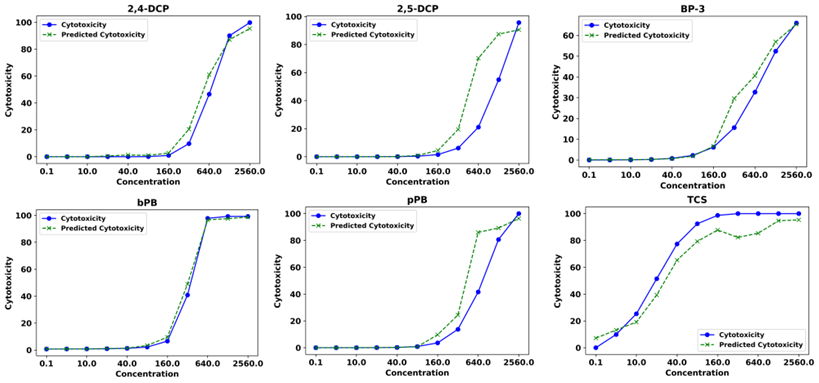}
\caption*{\textbf{Figure S5.} 4. Dose Response Curves for Fine-Tuned ME RF Model of Phenols.}
\end{figure}

\begin{figure}[!htbp]
\centering
\includegraphics[width=0.96\textwidth,keepaspectratio]{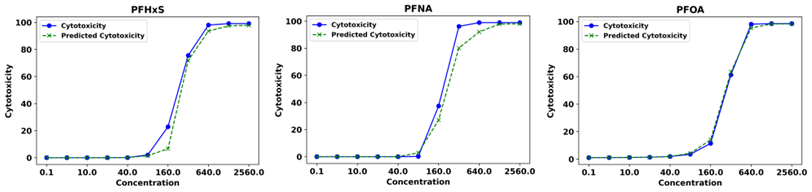}
\caption*{\textbf{Figure S5.} 3. Dose Response Curves for Fine-Tuned ME RF Model of Polyfluoroalkyl Compounds.}
\end{figure}

\begin{figure}[!htbp]
\centering
\includegraphics[height=0.78\textheight,keepaspectratio]{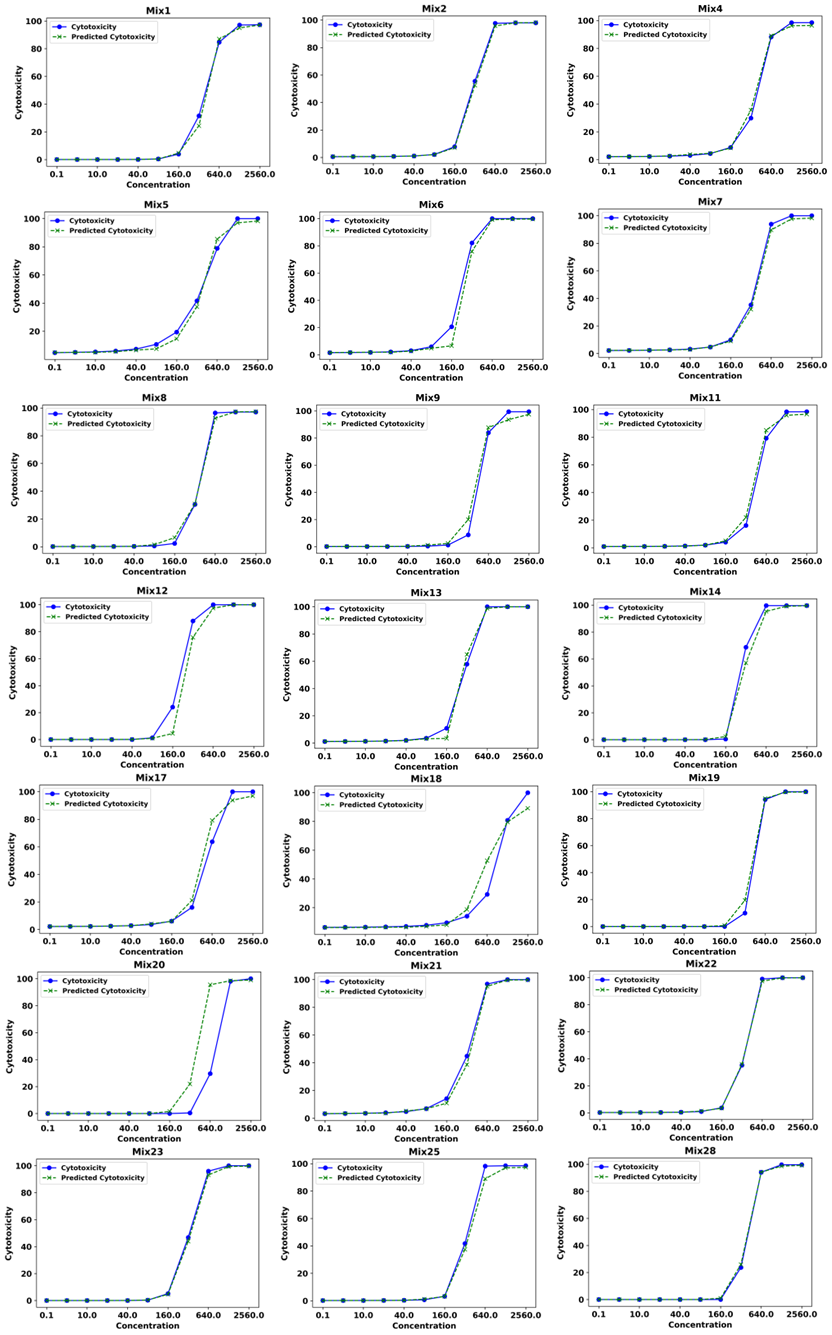}
\caption*{\textbf{Figure S5.} 5. Dose Response Curves for Fine-Tuned ME RF Model of Mixtures 1 to 28.}
\end{figure}

\begin{figure}[!htbp]
\centering
\includegraphics[height=0.78\textheight,keepaspectratio]{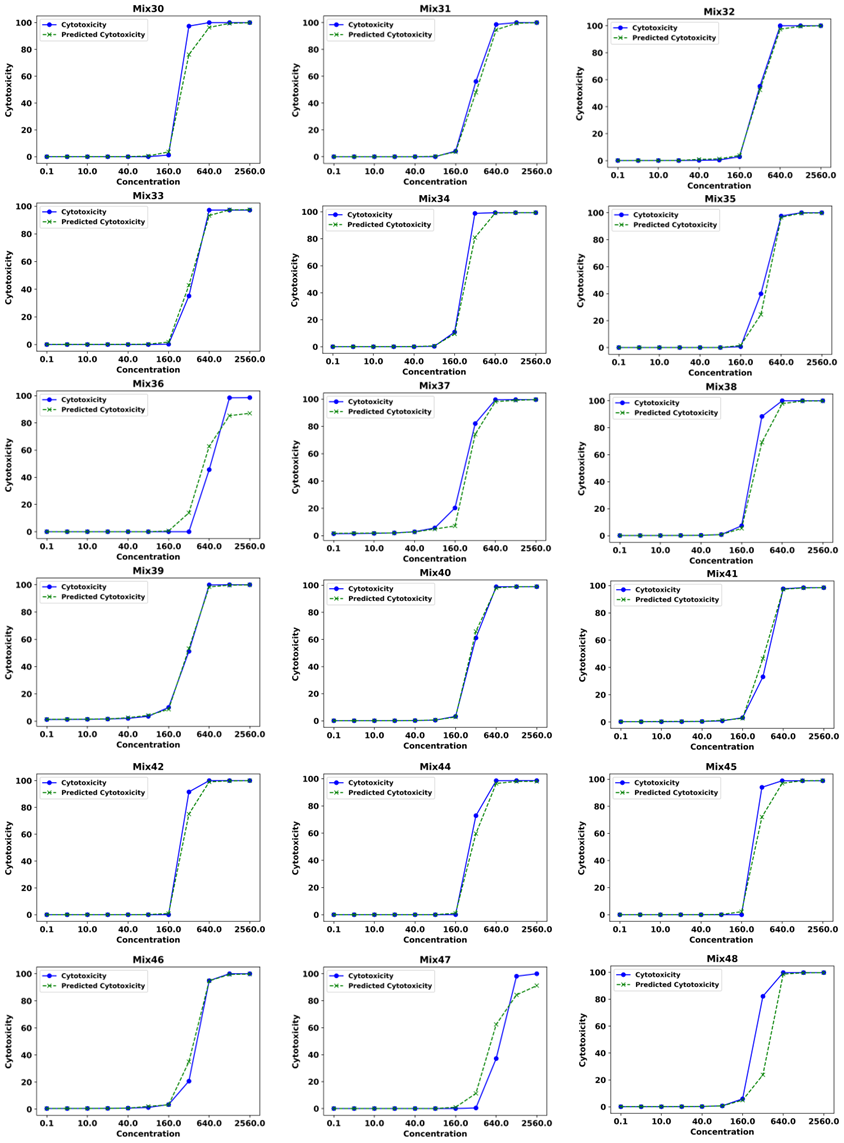}
\caption*{\textbf{Figure S5.} 6. Dose Response Curves for Fine-Tuned ME RF Model of Mixtures 30 to 48.}
\end{figure}

\clearpage